\documentclass[conference]{IEEEtran}
\IEEEoverridecommandlockouts
\usepackage{cite}
\usepackage{amsmath,amssymb,amsfonts}
\usepackage{algorithmic}
\usepackage{graphicx}
\usepackage{textcomp}
\usepackage{xcolor}

\usepackage{subcaption}
\usepackage{tikz}
\usetikzlibrary{quantikz2}
\usepackage{adjustbox}
\usepackage{algorithm}
\usepackage{algorithmic}
\usepackage{booktabs}
\usepackage{comment}

\def\BibTeX{{\rm B\kern-.05em{\sc i\kern-.025em b}\kern-.08em
    T\kern-.1667em\lower.7ex\hbox{E}\kern-.125emX}}

\begin{document}

\title{Evolutionary-Based Circuit Optimization for Distributed Quantum Computing
\thanks{© 2025 IEEE.  Personal use of this material is permitted.  Permission from IEEE must be obtained for all other uses, in any current or future media, including reprinting/republishing this material for advertising or promotional purposes, creating new collective works, for resale or redistribution to servers or lists, or reuse of any copyrighted component of this work in other works.

Sponsored in part by the Bavarian Ministry of Economic Affairs, Regional Development and Energy as part of the 6GQT project, as well as the German Federal Ministry of Research, Technology and Space under the funding code 01MQ22008A. The sole responsibility for the report's contents lies with the authors.}
}

\makeatletter
\newcommand{\linebreakand}{%
  \end{@IEEEauthorhalign}
  \hfill\mbox{}\par
  \mbox{}\hfill\begin{@IEEEauthorhalign}
}
\makeatother

\author{\IEEEauthorblockN{Leo Sünkel}
\IEEEauthorblockA{\textit{Institute for Informatics} \\
\textit{LMU Munich}\\
Munich, Germany \\
leo.suenkel@ifi.lmu.de}
\and
\IEEEauthorblockN{Jonas Stein}
\IEEEauthorblockA{\textit{Institute for Informatics} \\
\textit{LMU Munich}\\
Munich, Germany}
\and
\IEEEauthorblockN{Gerhard Stenzel}
\IEEEauthorblockA{\textit{Institute for Informatics} \\
\textit{LMU Munich}\\
Munich, Germany} 
\linebreakand
\IEEEauthorblockN{Michael Kölle}
\IEEEauthorblockA{\textit{Institute for Informatics} \\
\textit{LMU Munich}\\
Munich, Germany}
\and
\IEEEauthorblockN{Thomas Gabor}
\IEEEauthorblockA{\textit{Institute for Informatics} \\
\textit{LMU Munich}\\
Munich, Germany}
\and
\IEEEauthorblockN{Claudia Linnhoff-Popien}
\IEEEauthorblockA{\textit{Institute for Informatics} \\
\textit{LMU Munich}\\
Munich, Germany}
}

\maketitle

\begin{abstract}
In this work, we evaluate an evolutionary algorithm (EA) to optimize a given circuit in such a way that it reduces the required communication when executed in the Distributed Quantum Computing (DQC) paradigm. We evaluate our approach for a state preparation task using Grover circuits and show that it is able to reduce the required global gates by more than 89\% while still achieving high fidelity as well as the ability to extract the correct solution to the given problem. We also apply the approach to reduce circuit depth and number of CX gates. Additionally, we run experiments in which a circuit is optimized for a given network topology after each qubit has been assigned to specific nodes in the network. In these experiments, the algorithm is able to reduce the communication cost (i.e., number of hops between QPUs) by up to 19\%.
\end{abstract}


\section{Introduction}\label{sec:intro}
Constructing and optimizing quantum circuits through heuristic approaches like evolutionary algorithms (EAs) or machine learning based techniques, for instance, on the basis of reinforcement learning (RL), has become a popular research avenue in recent years \cite{ge2024quantum}. Exemplary tasks include optimizing existing circuits in order to reduce the depth \cite{fosel2021quantum}, number of gates \cite{kliuchnikov2013optimization} or CX gates \cite{gheorghiu2022reducing}; however, automating the design of completely novel circuits is also an active field of research \cite{sunkel2023ga4qco,sunkel2025quantum}. As quantum computers continue to mature, the need for appropriate software also grows, and thus efficient and flexible techniques are required. EAs are meta-heuristic algorithms that can be applied to a variety of problems spanning various domains.
Furthermore, quantum computers, or QPUs, must also scale to vastly larger dimensions than currently possible in order to achieve a quantum advantage or enable error-correction. In recent years, the notion of connecting QPUs through classical and quantum channels to create quantum networks to overcome the scaling problem of the monolithic approach is receiving increasing interest as such a framework would allow for a so-called distributed quantum computing (DQC) approach in which large quantum circuits are executed by a number of smaller connected QPUs \cite{denchev2008distributed,caleffi2024distributed,cuomo2020towards}. However, this technique introduces a communication overhead, especially when performed over a large quantum network. In such a network, entanglement is the crucial resource as it is required for communication protocols such as teleportation or remote CX gate execution. Thus, optimizing communication becomes a crucial task. Various approaches have been proposed and discussed in the literature regarding the problem of communication cost minimization in DQC; however, most works focus on optimizing the schedule or qubit assignment problem while not optimizing the circuit itself. We aim to address this issue in this paper, that is, we consider the problem of minimizing the communication between QPUs in the form of global gates, i.e., remote operations, by optimizing and refining the circuit itself rather than the qubit schedule. We apply an EA to this problem and evaluate it with three different fitness functions. We additionally run experiments where a circuit is optimized for a specific network architecture where the distance, i.e., number of hops between QPUs is taken into account during the fitness evaluation. Thus, the method proposed can be seen as a pre-processing or optimization step in a compilation stack for DQC where a circuit is optimized to be executed on a given quantum network.

\section{Background and Problem Formulation}\label{sec:background}
In this section, we introduce and define what we understand as quantum circuit optimization. We also give a brief overview of DQC, focusing on the important issues relevant to this work.

\subsection{Quantum Circuit Optimization}
For our purposes, we define quantum circuit optimization (QCO) as the process of adjusting and refining a given circuit in such a way that it maximizes (or minimizes) some objective function. This includes, for example, minimizing the depth, size, or multi-qubit gates of a given circuit. However, in order to ensure similarity or equivalence to the original circuit other constraints or metrics must be considered, turning this into a multi-objective optimization problem. A related problem is circuit synthesis, in which a circuit is constructed from scratch to fulfill some task. An overview of this field is given in \cite{ge2024quantum}. 

The authors define two objectives, and the following is the most relevant for this work \cite{ge2024quantum}:

\begin{equation}
    \max \sum_i |\bra{\psi_i} \hat{U} \ket{\chi_i}|^2
\end{equation}

\noindent where $\hat{U}$ is the constructed unitary and $\chi$ and $\psi$ are the initial and final states respectively.
However, in this work, we solely focus on QCO, that is, we start with a given circuit and optimize it through an EA to maximize some objective function. Moreover, we focus on state preparation; that is, we start in a defined initial state, and thus the fidelity objective becomes: 

\begin{equation}
    \max |\bra{\psi} \hat{U} \ket{0}|^2
\end{equation}

\noindent where $\ket{0}$ signifies that we start with all qubits in this state, i.e., this is the initial state and $\bra{\psi}$ represents the target state. However, in our approach, we adjust the circuit to fulfill the given fidelity as well as other constraints, which we define in Section \ref{sec:approach}.

\subsection{Distributed Quantum Computing}
DQC \cite{caleffi2024distributed} provides a pathway to scale quantum computing to dimensions that are difficult, if not impossible, to achieve with the current monolithic approach. By connecting multiple QPUs one can enable DQC to execute circuits that are too large for a single machine. DQC can be achieved via QPUs arranged in a cluster, i.e., directly connected to each other, or alternatively, QPUs can be connected through a large-scale quantum communication network. However, the latter approach would require more sophisticated hardware and algorithms to enable long-distance communication with high fidelity. For example, quantum repeaters \cite{briegel1998quantum} enabling entanglement swapping and purification are crucial for accomplishing this task. 
In such a network, entanglement and Bell pairs in particular are the central resource as they enable quantum communication through teleportation, an important protocol in DQC. Before a multi-qubit gate such as a CX (i.e., controlled not) gate where the control and target qubits are located in different QPUs can be executed, the qubits must first be teleported such that they are located in the same QPU in order for the gate to be performed locally. The teleportation protocol is shown in Fig. \ref{fig:teleportation}. Alternatively to teleporting the qubits, a remote CX can be performed. Both protocols consume the same resources, and which is better depends on the specific circuit and constraints, i.e., in some situations it might be favorable to teleport specific qubits while in other situations a remote CX is preferred in order to minimize the overall communication. Thus, the qubit scheduling problem, i.e., assigning qubits to available QPUs, becomes a crucial optimization problem and is part of the main topic of this work. Various approaches and strategies have been proposed by the community, and we will give a brief overview in the next section; however, we will first discuss the problem formulation.

\begin{figure}[tb]
    \centering
    \boxed{
        \begin{quantikz}
            \lstick{\ket{\psi}} & \ctrl{1} & \gate{H} & \meter{} \wire[d][2]{c} \\
                \lstick[2]{\ket{\Phi_+}} & \targ{} &  & & \meter{} \wire[d][1]{c} \\
                 & & & \gate{X} & \gate{Z} & \ket{\psi} \\
        \end{quantikz}
        }
    \caption{The teleportation protocol \cite{nielsen2010quantum}. The state $\ket{\psi}$ is teleported from the top qubit to the qubit at the bottom. Note that the state is no longer present at its original location.}
    \label{fig:teleportation}
\end{figure}
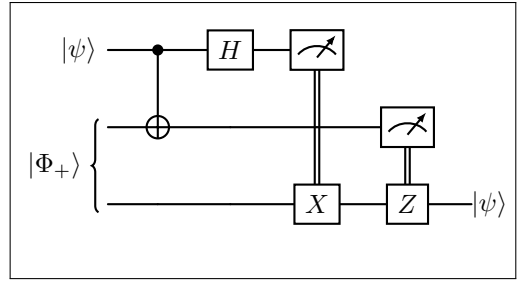

\subsection{Problem Formulation}
The overall objective of the EA is to optimize a given circuit so that its communication is minimized when executed in a distributed manner. As discussed above, communication between QPUs occurs when a multi-qubit gate is executed where the involved qubits are located in different QPUs. Multi-qubit gates where the control and target qubits are located in different QPUs are referred to as global gates, and the only multi-qubit gates considered in this work are CX-gates. The objective is thus to assign qubits to QPUs such that global gates are minimized. More formally, this objective can be defined as follows:

\begin{equation}
    c(p, cx) = \sum_{(i, j) \in cx} 1 - \delta(p(i), p(j))
\end{equation}

\begin{equation}
    \min_p c(p, cx)
\end{equation}

\noindent where $p$ is the partitioning, that is, the assignment of qubits to QPUs (e.g., as integers), and $cx$ a list of all CX-gates (i.e., the indices of the control and target qubits) in a particular circuit, $\delta$ denotes the Kronecker delta function.
Alternatively, when the topology of the underlying network of QPUs should be considered, the fitness function can be adjusted to incorporate the distance between the nodes:

\begin{equation}\label{eq:distance}
    g(p, cx) = \sum_{(i, j) \in cx} \textnormal{dist}(p(i), p(j)) 
\end{equation}

\noindent where $\textnormal{dist}$ is the number of hops between the QPUs.

A common approach is to treat this problem as a scheduling or assignment problem in which qubits are assigned to different QPUs; however, the circuit itself is not optimized. In this paper, we suggest optimizing the circuit so that it is more suitable for execution in a DQC environment. In addition to minimizing the communication cost, the state produced by the circuit must remain as similar as possible to the target state. Thus, the fidelity between the target state $\ket{t}$ and the state prepared by the circuit $U$ must be maximized:

\begin{equation}
    F(|t\rangle, U) = |\langle t|U|0\rangle|^2
\end{equation}

The objective can then be defined as follows:

\begin{equation}\label{eq:objective}
    f(|t\rangle, p, U) = \alpha \cdot F(|t\rangle, U) - g(p, U_{cx})
\end{equation}

\begin{equation}
    \max_U  f(|t\rangle, p, U)
\end{equation}

\noindent where $\alpha$ is a weighting factor for fidelity, $|t\rangle$ is the target state, $p$ a partitioning, $U$ the circuit corresponding to a particular solution, and $U_{cx}$ all CX gates present in a given circuit. Note that $g$ is rescaled with respect to the original circuit (with the original circuit corresponding to a value of 1). The partitioning $p$ can be determined by various algorithms, such as graph partitioning, for example, and can be fixed or computed dynamically in the fitness calculation. We give more details on how $p$ is determined in this work in the experimental setup section below.

\section{Related Work}\label{sec:related_work}
The problem of minimizing communication costs, and the number of required teleportations in particular, has been receiving increasing attention by the research community working in the DQC field. Various evolutionary and genetic algorithms have been applied to this task \cite{houshmand2020evolutionary,zomorodi2018optimizing,sunkel2024applying,crampton2024genetic,burt2024generalised} as well as QUBO-based methods \cite{chen2024circuit,bandic2023mapping}. However, these approaches tend to focus on optimizing the schedule or qubit assignment, and not the circuit itself. In \cite{sunkel2025time}, the authors evaluate different approaches to minimize the communication cost, including optimizing the circuit itself through an EA, which is a similar approach to the one discussed in this work. Compilation for DQC has been addressed, for example, in \cite{cuomo2023optimized,ferrari2021compiler} and a survey on the field of DQC is given in \cite{caleffi2024distributed}.
The construction and optimization of quantum circuits have also been extensively studied in recent years, most notably approaches based on EAs \cite{lukac2002evolving,lukac2003evolutionary,potovcek2018multi,ding2022evolutionary,sunkel2023ga4qco,ding2023multi} and RL \cite{kuo2021quantum,fosel2021quantum,ostaszewski2021reinforcement,kolle2024reinforcement}.
The combination of both fields to the problem of minimizing the communication cost in DQC is the main topic and contribution of this work, and we introduce our approach next.

\section{Approach}\label{sec:approach}
The problem formulation of the algorithm and the evolutionary operations are similar to the EA proposed in \cite{sunkel2025time}.
\subsection{Evolutionary Algorithm}
\subsubsection{Encoding}
Circuits are encoded using a one-dimensional list where each element represents a gate, and each gate is an object containing all relevant information such as name, gate type, qubits and parameters. An example of this encoding is shown in Fig. \ref{fig:example_encoding}. Based on this representation, a Qiskit circuit is constructed, which can then be executed or used for other tasks.

\begin{figure}[tb]
    \centering
    \begin{subfigure}[b]{0.45\textwidth} 
        \centering
        \includegraphics[]{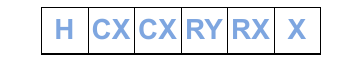}
        \caption{Example solution encoding.}
        \label{fig:enter-label}
    \end{subfigure}
    \hfill
    \begin{subfigure}[b]{0.45\textwidth} 
        \centering
        \begin{quantikz}
            \lstick{\ket{\psi_0}} & & \targ{} & \gate{RX} & \\
            \lstick{\ket{\psi_1}} & \ctrl{1} & & \gate{RY} &\\
            \lstick{\ket{\psi_2}} & \targ{} & & \gate{X} & \\
            \lstick{\ket{\psi_3}} & \gate{H} & \ctrl{-3} & &
        \end{quantikz}
        \caption{Corresponding circuit.}
    \end{subfigure}
    \caption{Example solution encoding with corresponding circuit. Note that in the encoding, we show the gate names; however, each gate object contains all relevant information (e.g., qubits, parameters, etc.) and thus the above circuit example is just one possible manifestation of that encoding.}
    \label{fig:example_encoding}
\end{figure}

\subsubsection{Crossover}
The algorithm utilizes two crossover methods, namely single-point and uniform crossover, which are illustrated in Fig. \ref{fig:spc} and Fig. \ref{fig:uc} respectively. In the first crossover method, a random cutoff point is determined and a new solution is created by taking all genes left to this point from one parent and all genes right of the cutoff point from another parent. With uniform crossover, a child is created by taking a single gene from either parent with a probability of 50\%. As solutions can have different lengths, the length of the child is also randomly inherited from either parent with equal probability. In the case of the child inheriting the larger circuit, the remaining genes are taken from the larger parent.

\begin{figure}[tb]
    \centering
    \begin{subfigure}[t]{0.45\textwidth}
        \centering
        \includegraphics[scale=0.5]{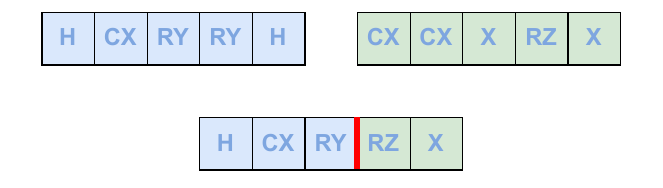}
        \caption{Single point crossover. The crossover point is marked red.}
        \label{fig:spc}
    \end{subfigure}
    \hfill
    \begin{subfigure}[t]{0.45\textwidth}
        \centering
        \includegraphics[scale=0.5]{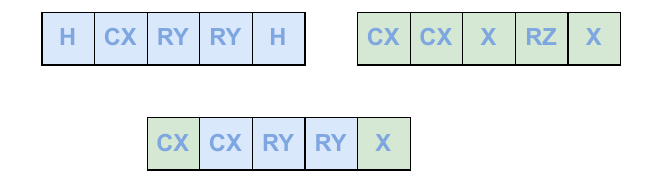}
        \caption{Uniform crossover}
        \label{fig:uc}
    \end{subfigure}
    \caption{The EA utilizes two different crossover methods: single point and uniform crossover. Which one is used is determined randomly.}
\end{figure}

\subsubsection{Mutation}
Solutions can be adjusted by applying a mutation operation. We implemented the following six mutation methods: (1) gate flip, (2) delete gate, (3) swap gate, (4) shuffle, (5) add gate, and (6) remove cx gate. Their functioning is shown in Fig. \ref{fig:mutations_overview}.
The mutation method applied is determined randomly, and each individual is mutated only once.

\begin{figure}[tb]
    \centering
    \begin{subfigure}[t]{0.9\linewidth}
        \centering
        \includegraphics[width=\linewidth]{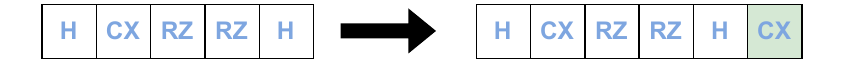}
        \caption{Add gate mutation}
    \end{subfigure}
    
    \begin{subfigure}[t]{0.9\linewidth}
        \centering
        \includegraphics[width=\linewidth]{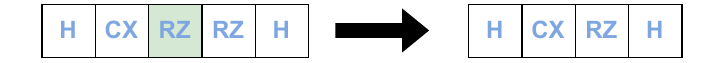}
        \caption{Delete gate mutation}
    \end{subfigure}
    
    \begin{subfigure}[t]{0.9\linewidth}
        \centering
        \includegraphics[width=\linewidth]{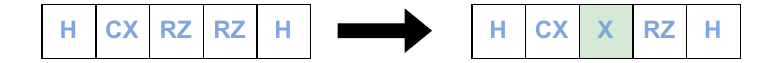}
        \caption{Gate flip mutation}
    \end{subfigure}
    
    \begin{subfigure}[t]{0.9\linewidth}
        \centering
        \includegraphics[width=\linewidth]{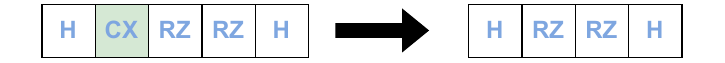}
        \caption{Remove CX gate mutation}
    \end{subfigure}
    
    \begin{subfigure}[t]{0.9\linewidth}
        \centering
        \includegraphics[width=\linewidth]{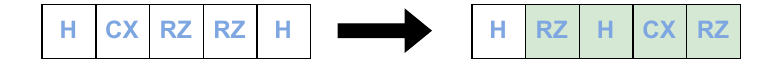}
        \caption{Shuffle mutation}
    \end{subfigure}
    
    \begin{subfigure}[t]{0.9\linewidth}
        \centering
        \includegraphics[width=\linewidth]{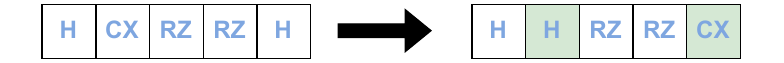}
        \caption{Swap gates mutation}
    \end{subfigure}
    
    \caption{Overview of all mutation operations implemented.}
    \label{fig:mutations_overview}
\end{figure}

\subsubsection{Evolution}
Individuals are initialized with the target circuit as the solution, and this forms the initial population. The algorithm then aims to optimize solutions according to the given fitness function. New individuals, i.e., solutions, are created through either crossover or new initializations. A random mutation is then applied to each solution according to a set probability. The worst $n$ solutions are replaced by the best $n$ ``children'', i.e., new solutions. The child rate parameter specifies how many children are created in each generation, whereas the  ``replace rate'' how many solutions of the population should be replaced. The parameters used in the experiments are shown in Table \ref{tab:hyperparameters}. Note that the crossover rate specifies how many new individuals, i.e., the offspring, should be created by crossover, and how many will be initialized as new individuals without parents.

\subsection{Fitness}
The overall objective is to optimize a given circuit so that its communication cost, measured in the number of global gates, is minimized while still maintaining high fidelity to the original state prepared by the target circuit. 
While the objective defined in Equation \ref{eq:objective} seeks to directly minimize global gates, i.e., CX gates between different QPUs, other proxy fitness functions are also investigated in this work, including the minimization of circuit depth and total number of CX gates.
The minimize depth fitness function is defined as follows:

\begin{equation}
    f_{depth}(|t\rangle, U) = \alpha \cdot F(|t\rangle, U) - U_{depth}
\end{equation}

\begin{equation}
    \max_U f_{depth}(|t\rangle, U)
\end{equation}

\noindent where $U_{\textnormal{depth}}$ is the rescaled depth of the circuit $U$ with respect to the original circuit.
The fitness function to minimize the total number of CX gates is defined as

\begin{equation}
    f_{cx}(|t\rangle, U) = \alpha \cdot F(|t\rangle, U) - U_{cx}
\end{equation}

\begin{equation}
    \max_U f_{cx}(|t\rangle, U)
\end{equation}

\noindent where $U_{cx}$ is the rescaled total number of CX gates present in circuit $U$ with respect to the original circuit.

The distance, i.e., communication cost or number of global gates is calculated as specified in Equation \ref{eq:distance}. In experiments with 4, 5, and 6 qubits, we consider 2 partitions, i.e., a network consisting of 2 QPUs. In this case, the Kernighan-Lin algorithm \cite{kernighan1970efficient} is used to calculate the partitioning $p$ and thus the number of global gates. We also run experiments with 8 qubits, and here a network consisting of 4 nodes arranged in a grid where each QPU has a capacity of 2 qubits is used. In this scenario, a naive partitioning is given where each QPU is assigned 2 consecutive qubits, that is, QPU 0 is assigned qubits 0 and 1, QPU 1, 2 and 3, and so forth. Thus, in this experiment, the distance, i.e., the number of hops between the QPUs is taken into account during the fitness calculation and the circuit must be optimized accordingly. We ran an additional experiment with 6 qubits of this nature where a 3 node network topology was used.

\section{Experimental Setup}\label{sec:experimental_setup}
We evaluate the approach proposed on Grover circuits for preliminary results as part of this study. More specifically, we created circuits that aim to search for a given target state specified by a bitstring (e.g., 0110 for a 4 qubit implementation), and the target states were determined randomly. In the resulting distribution after the circuit is run, the target state should have the highest amplitude. The extracted solution resulting from the optimized circuit is compared to the original target to determine whether it is still able to produce the desired output. We created circuits containing 4, 5, 6 and 8 qubits, each with a random target, and the circuits were created using Qiskit. Note that the circuits were transpiled using the following basis gates: x, sx, rz, and cx, and the EA only uses these gates. We run the experiments with the different fitness functions discussed above and evaluate how each affects all of the specified metrics (i.e., depth, number of CX, and global gates (remote operations)).
In the experiments with the 8 qubit circuit, a network with 4 nodes was used and the distance, i.e., number of hops between nodes, was used in the fitness function. We also ran experiments with a 6 qubit circuit with a network consisting of 3 nodes. Note that each QPU has an equal capacity of 2 qubits. 
The hyperparameters used in the experiments are listed in Table \ref{tab:hyperparameters}, and experiments were run for 3 different seeds with the listed configuration. Qiskit \cite{qiskit2024} was used to implement and simulate the circuits as well as for calculating the fidelity. The Kernighan-Lin graph partitioning implementation as well as other graph functions were used from NetworkX \cite{SciPyProceedings_11}.

\begin{table}[tb]
    \centering
    \caption{Hyperparameters used were determined experimentally.}
    \label{tab:hyperparameters}
    \begin{tabular}{l@{\hspace{2cm}}r}
        \toprule
         Population size & 200  \\ \midrule
         Generations & 3000 \\ \midrule
         Crossover rate & 0.85 \\ \midrule
         Mutation rate & 0.4 \\ \midrule
         Child rate & 0.3 \\ \midrule
         Replace rate & 0.1 \\ \midrule
         Fidelity weight & 1 (2 and 3 in the network experiments)\\ 
        \bottomrule
    \end{tabular}  
\end{table}

\begin{figure*}[t]
    \centering
    \begin{subfigure}[b]{0.33\textwidth}
        \includegraphics[scale=0.3]{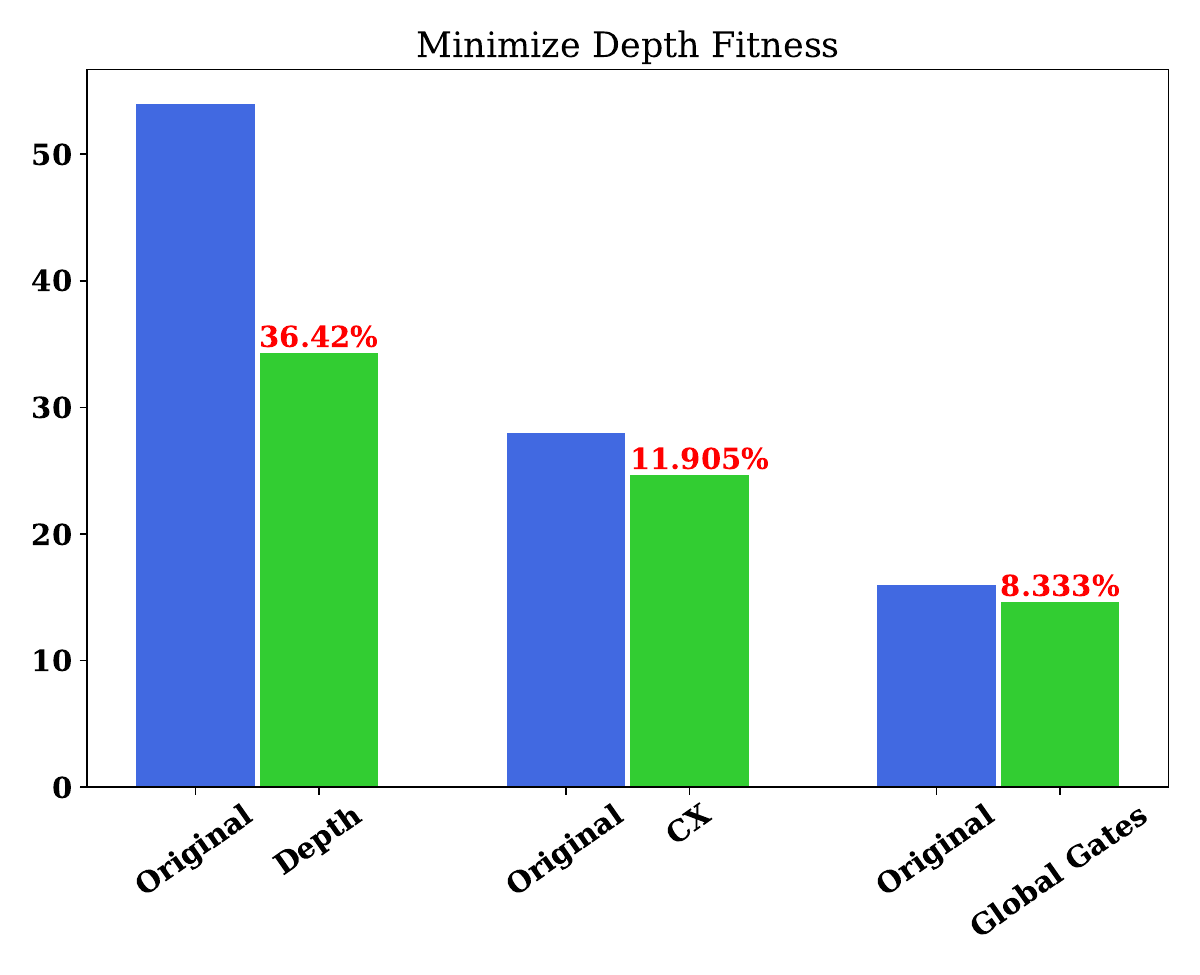}
        \caption{4 qubits}
    \end{subfigure}
    \hfill
    \begin{subfigure}[b]{0.33\textwidth}
        \includegraphics[scale=0.3]{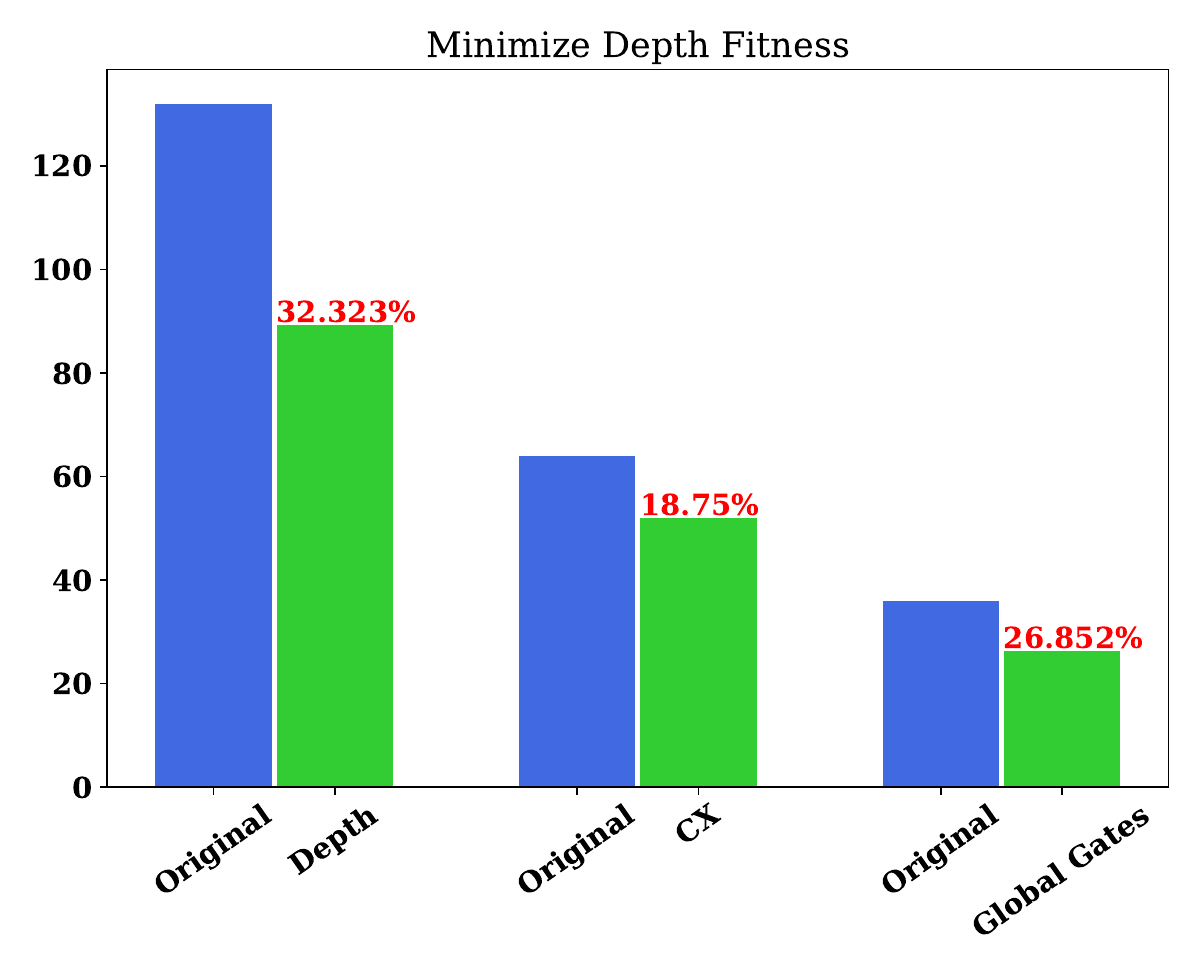}
        \caption{5 qubits}
    \end{subfigure}
    \hfill
    \begin{subfigure}[b]{0.32\textwidth}
        \includegraphics[scale=0.3]{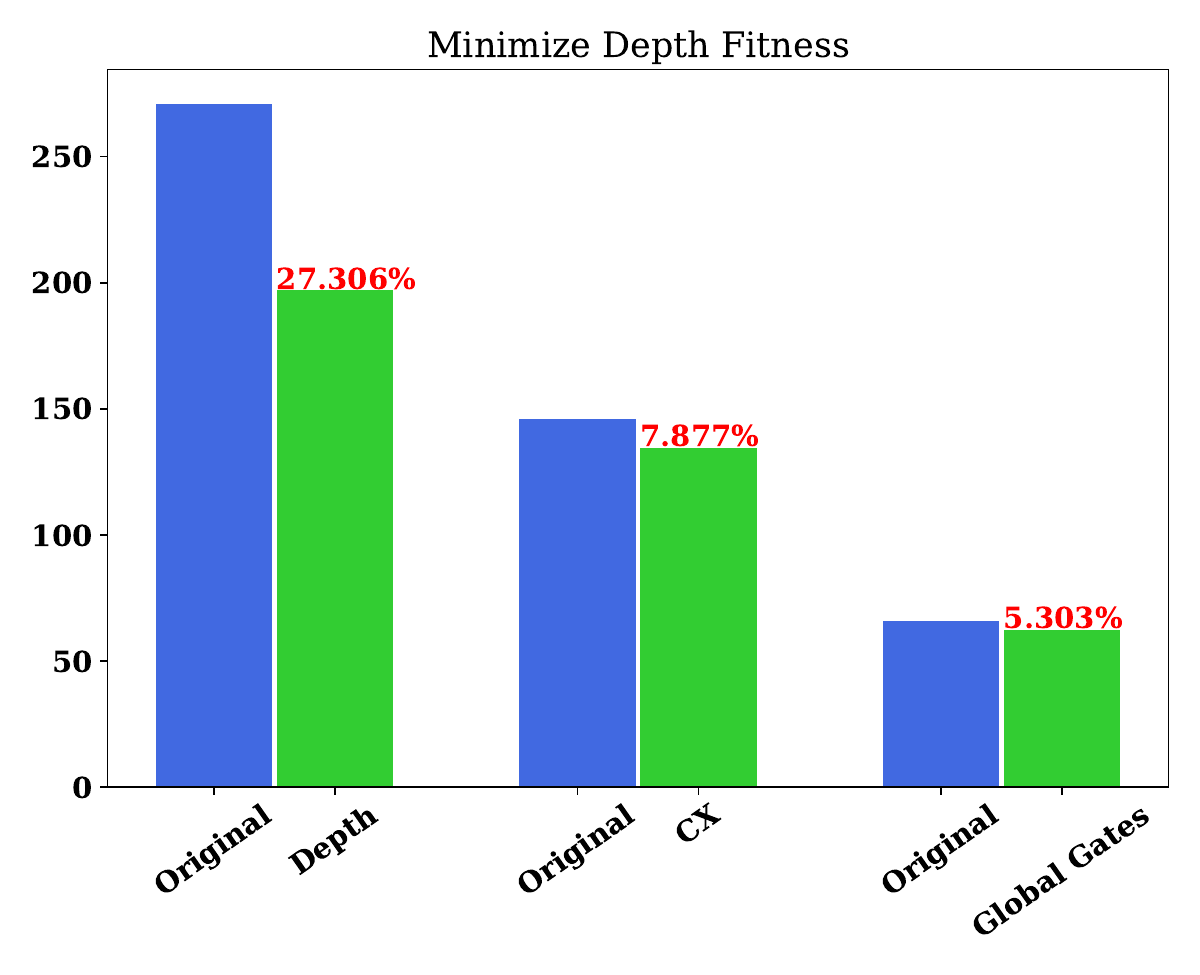}
        \caption{6 qubits}
    \end{subfigure}
    \caption{Results of the fitness function to reduce the depth of the circuit.}
    \label{fig:results_reduce_depth}

    \begin{subfigure}[b]{0.33\textwidth}
        \includegraphics[scale=0.3]{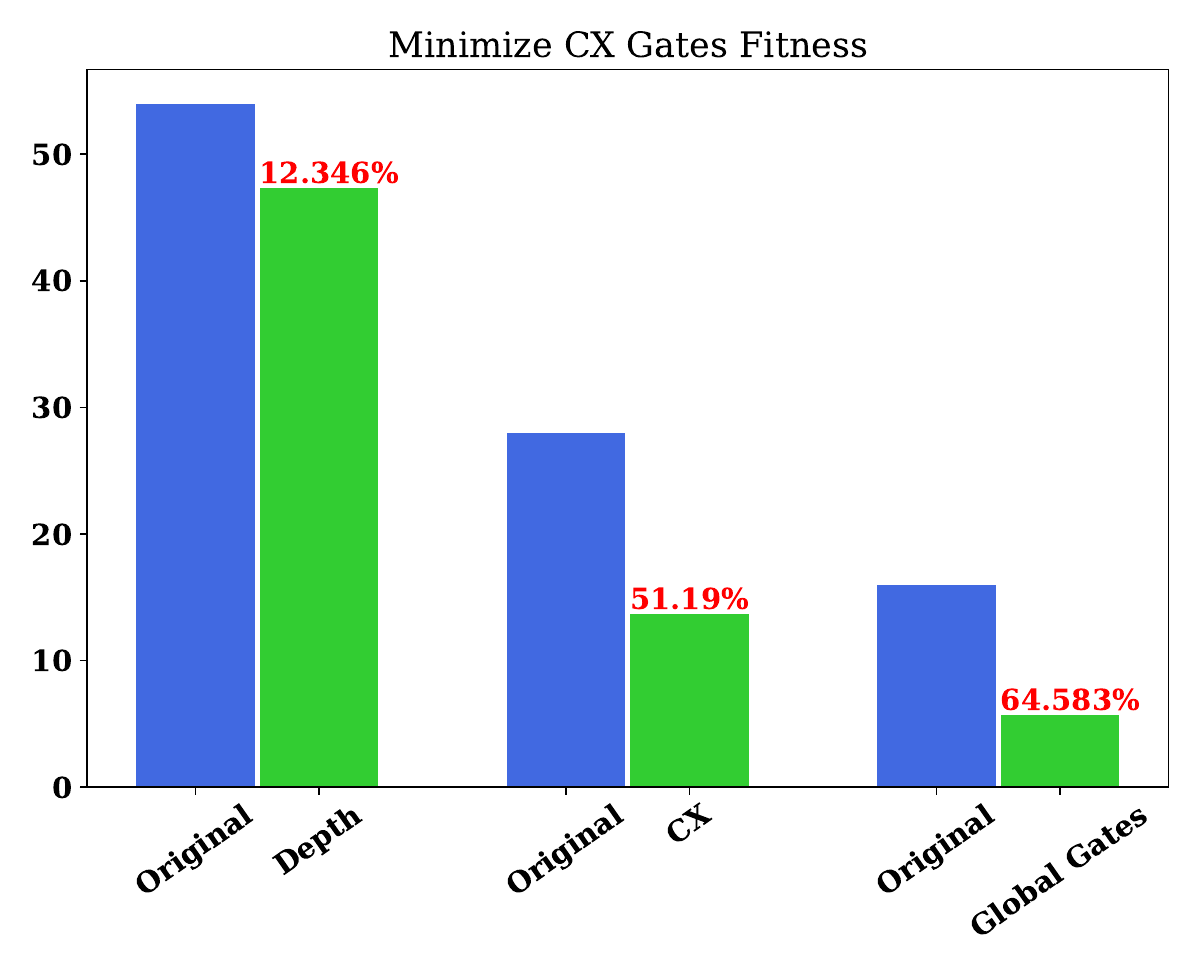}
        \caption{4 qubits}
    \end{subfigure}
    \hfill
    \begin{subfigure}[b]{0.33\textwidth}
        \includegraphics[scale=0.3]{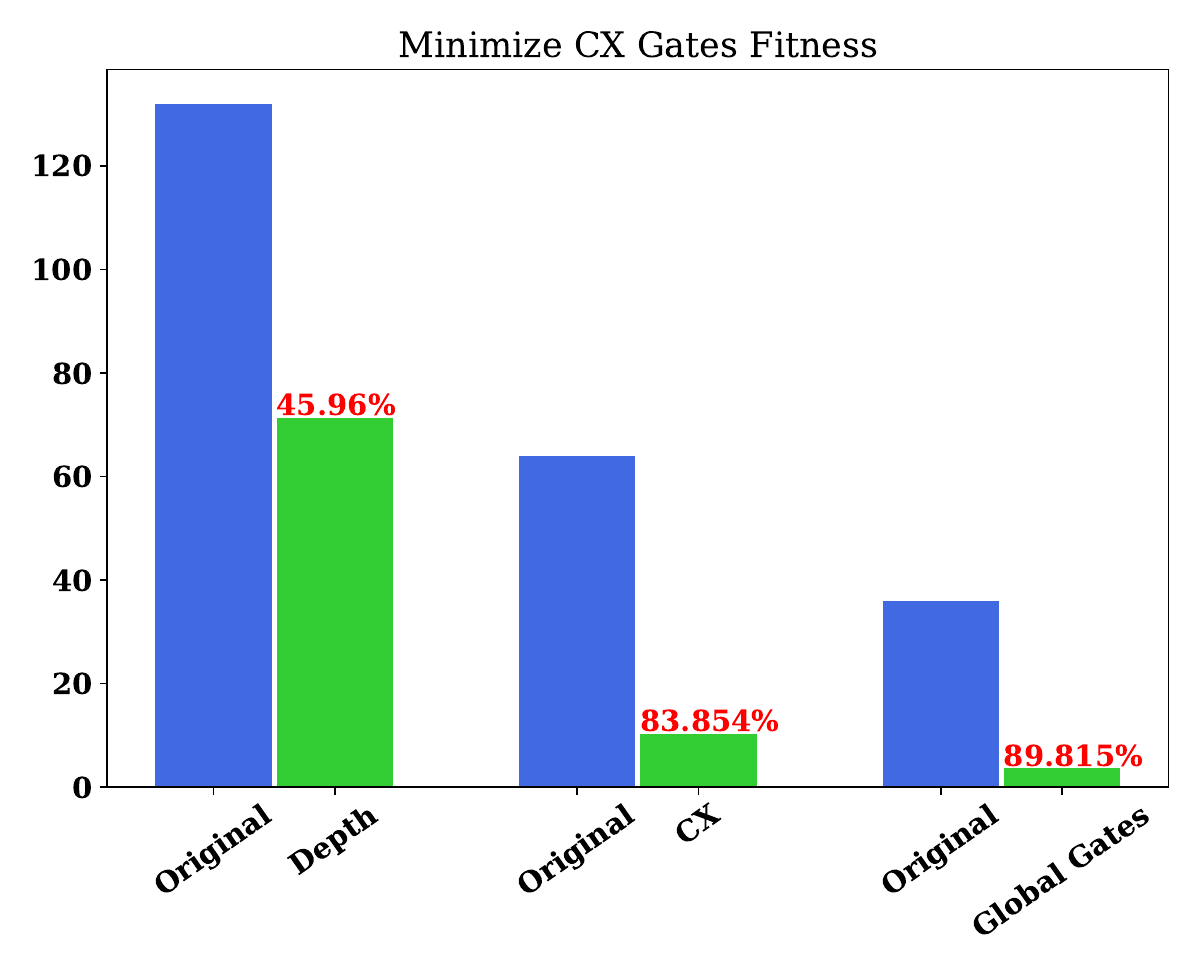}
        \caption{5 qubits}
    \end{subfigure}
    \hfill
    \begin{subfigure}[b]{0.32\textwidth}
        \includegraphics[scale=0.3]{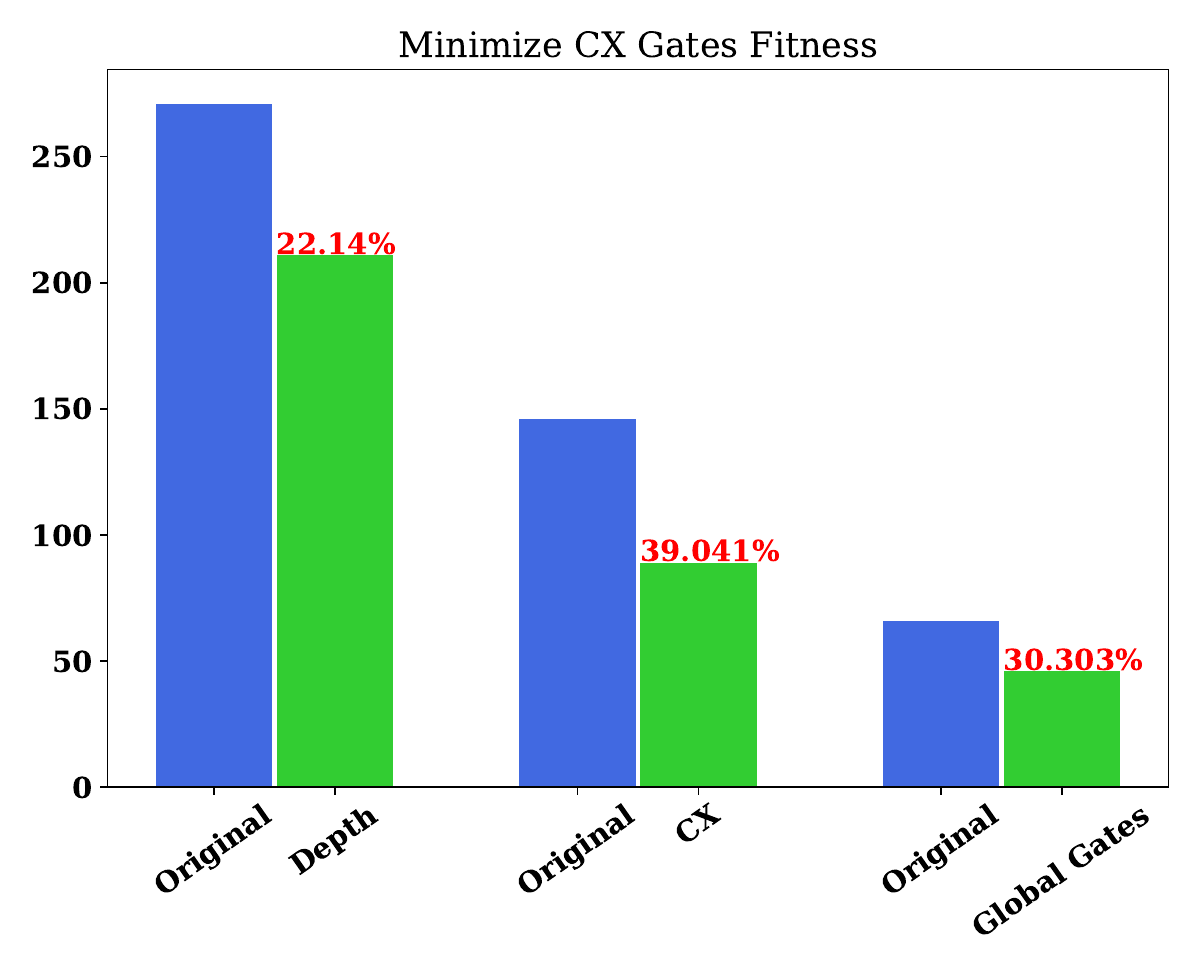}
        \caption{6 qubits}
    \end{subfigure}
    \caption{Results of the fitness function to reduce the total number of CX gates.}
    \label{fig:results_reduce_cx}

    \begin{subfigure}[b]{0.33\textwidth}
        \includegraphics[scale=0.3]{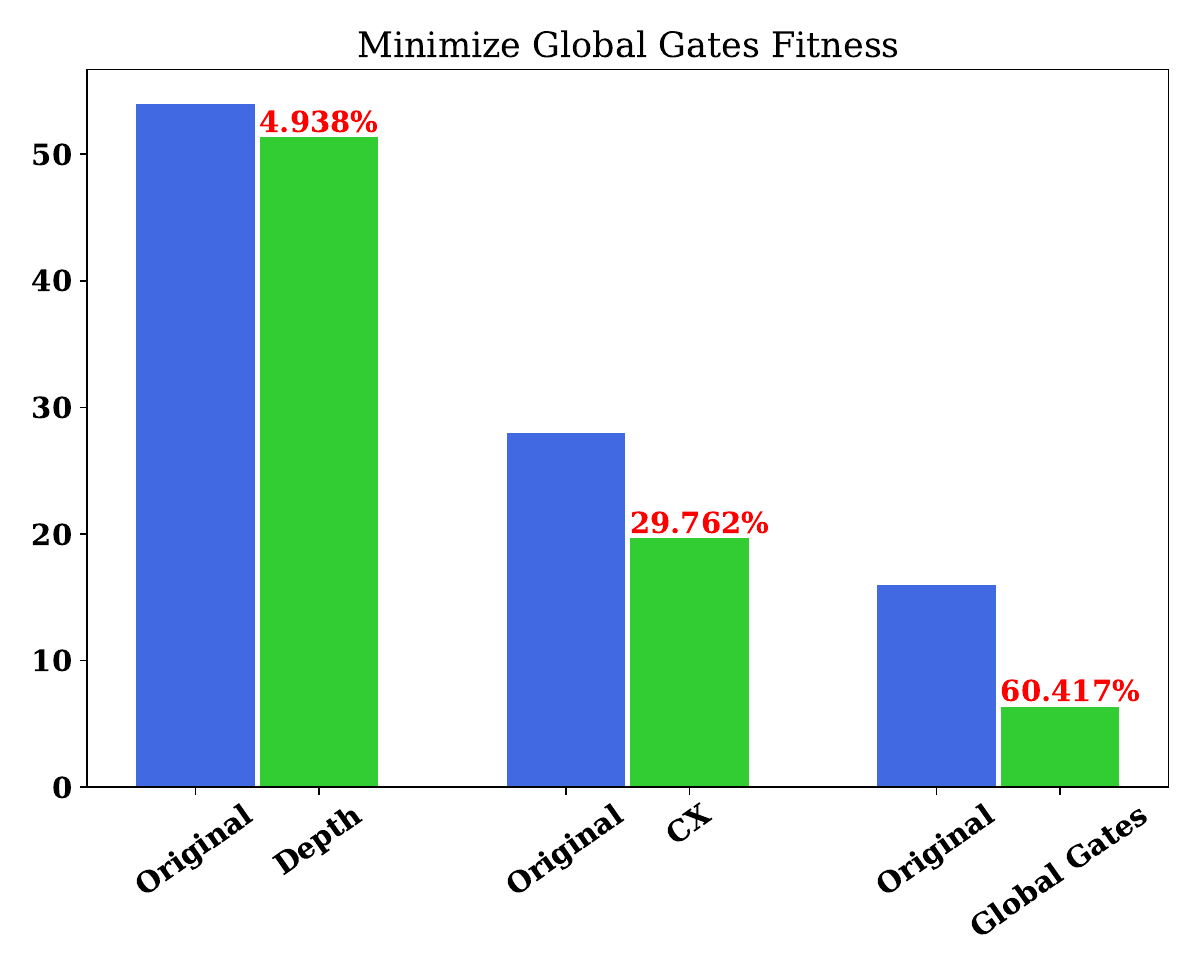}
        \caption{4 qubits}
    \end{subfigure}
    \hfill
    \begin{subfigure}[b]{0.33\textwidth}
        \includegraphics[scale=0.3]{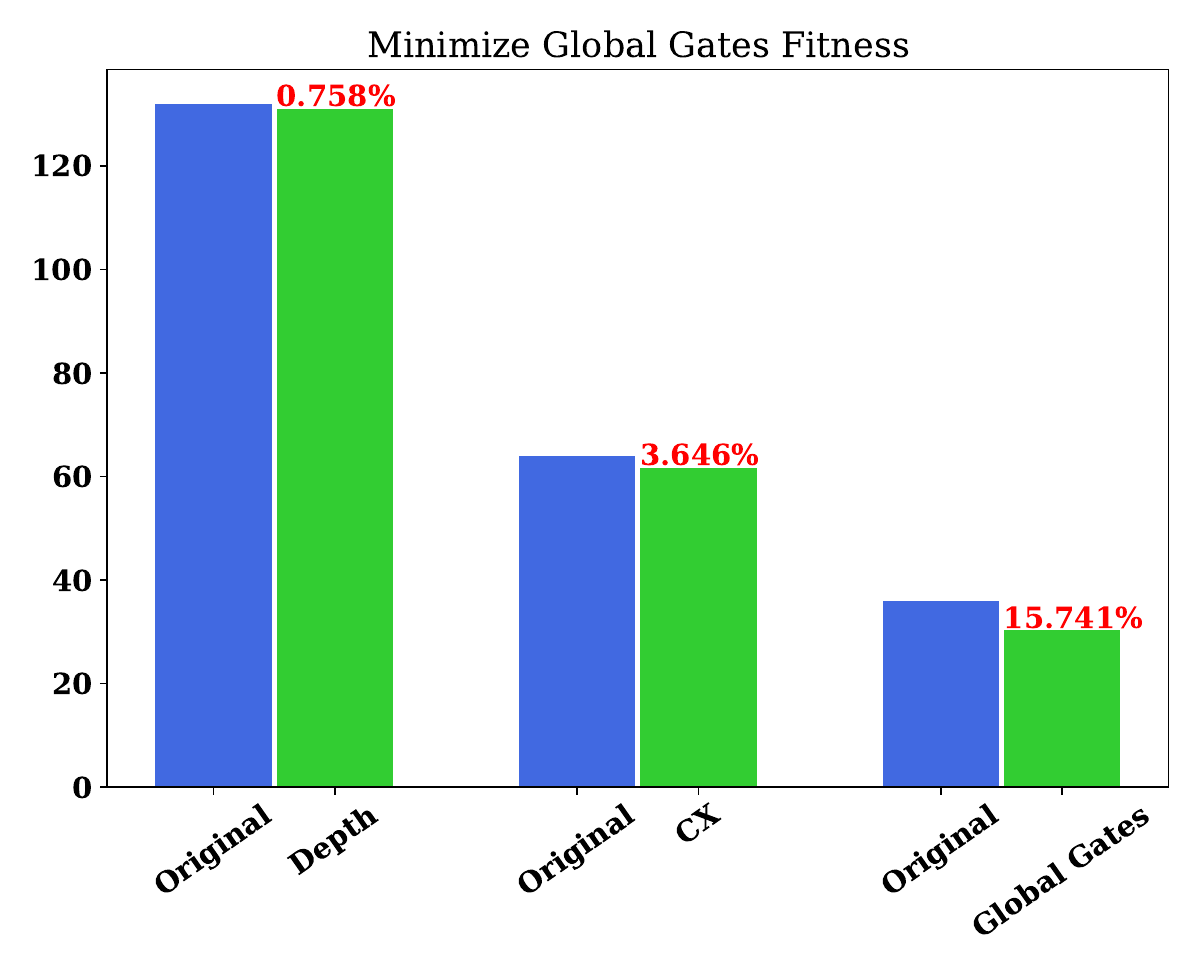}
        \caption{5 qubits}
    \end{subfigure}
    \hfill
    \begin{subfigure}[b]{0.32\textwidth}
        \includegraphics[scale=0.3]{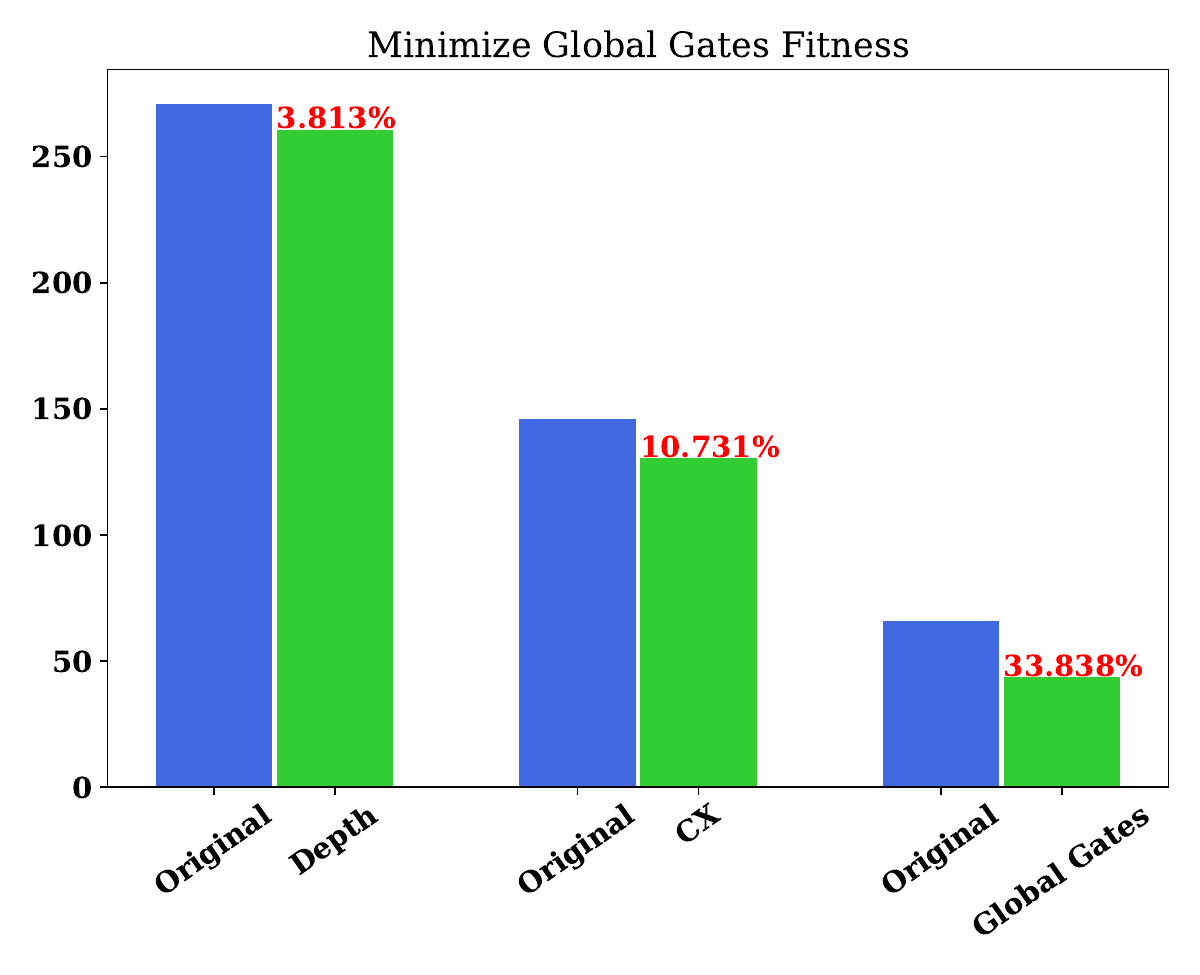}
        \caption{6 qubits}
    \end{subfigure}
    \caption{Results of the fitness function to reduce the number of global gates.}
    \label{fig:results_reduce_gg}
\end{figure*}

\begin{figure*}[tb]
    \centering
    \begin{subfigure}[b]{0.45\textwidth}
        \includegraphics[scale=0.5]{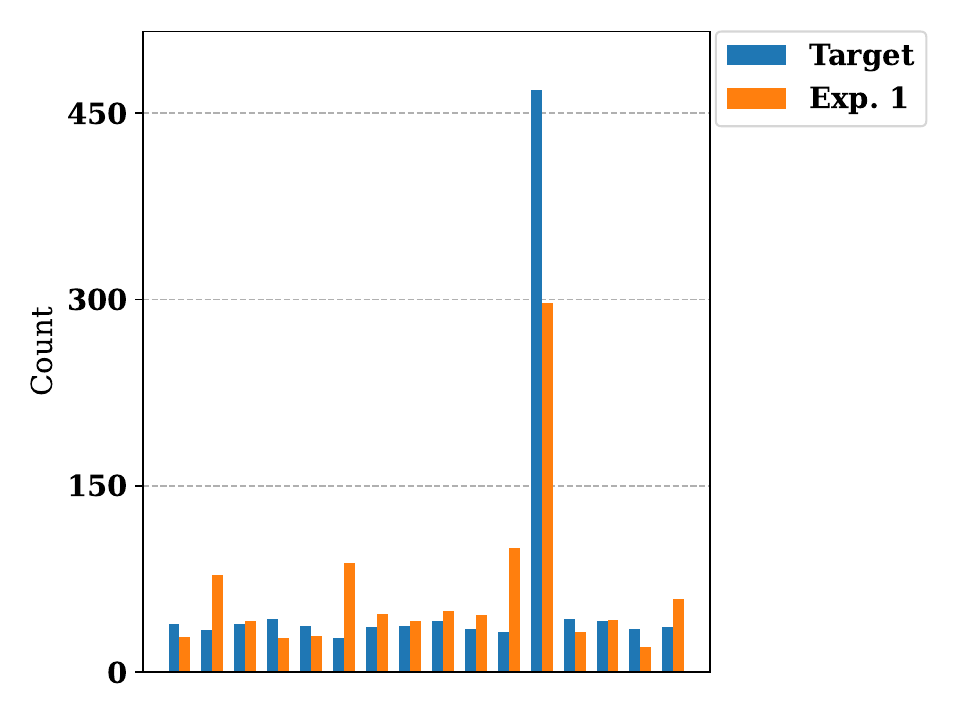}
        \caption{4 qubits distribution (fidelity weight 1).}
    \end{subfigure}
    \hfill
    \begin{subfigure}[b]{0.45\textwidth}
        \includegraphics[scale=0.5]{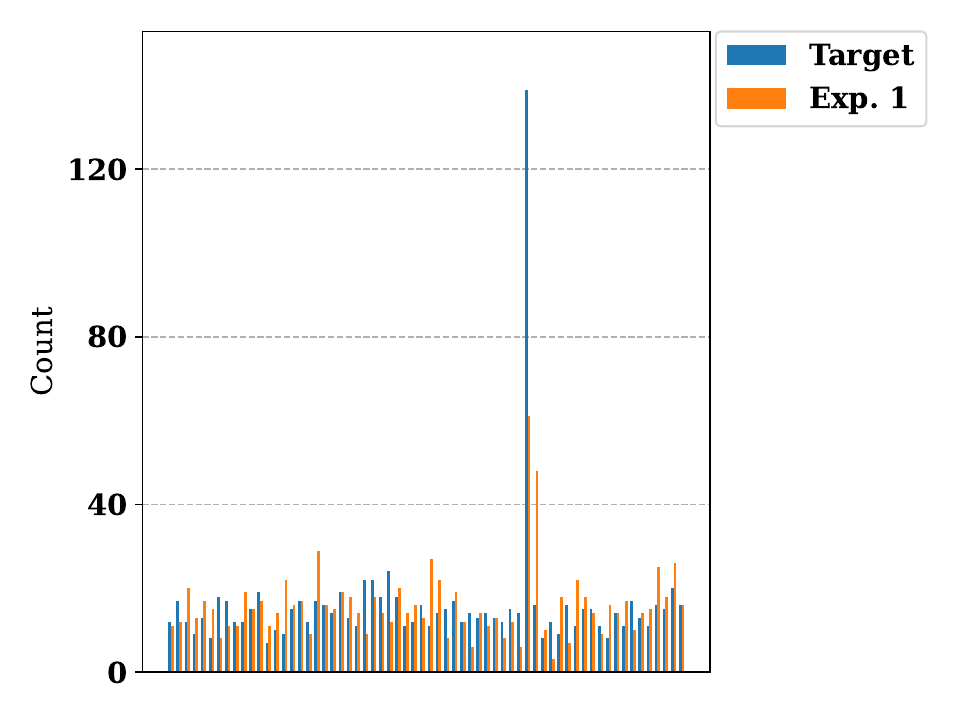}
        \caption{6 qubits distribution (fidelity weight 1).}
    \end{subfigure}
    \hfill
    \begin{subfigure}[b]{0.45\textwidth}
        \includegraphics[scale=0.5]{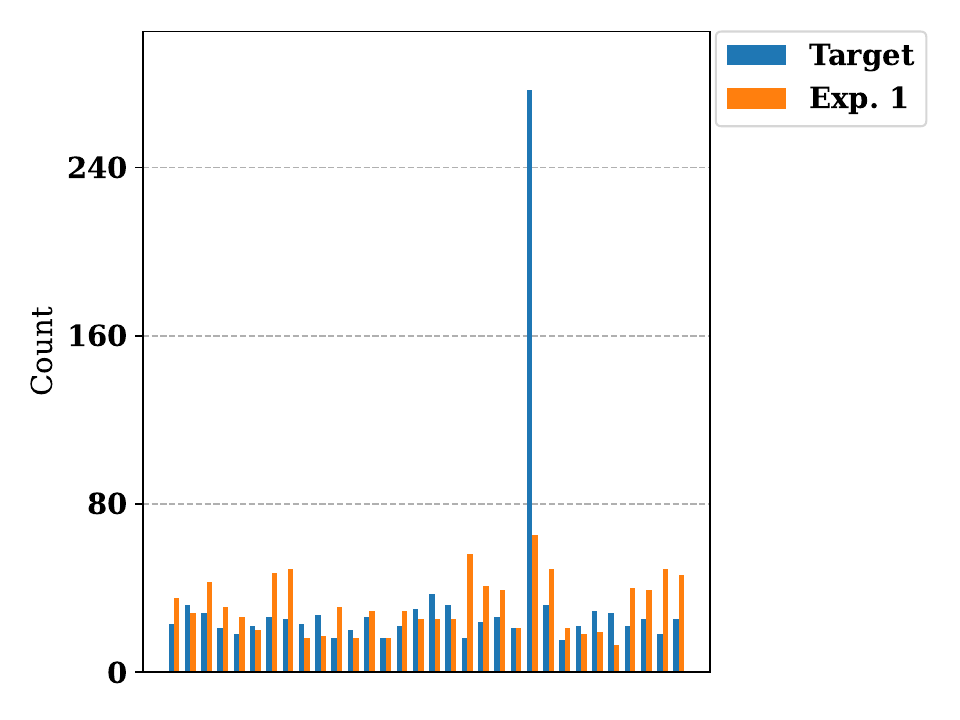}
        \caption{5 qubits distribution where the fidelity weight in the fitness is set to 1.}
        \label{fig:5_qubit_counts_weight1}
    \end{subfigure}
    \hfill
    \begin{subfigure}[b]{0.45\textwidth}
        \includegraphics[scale=0.5]{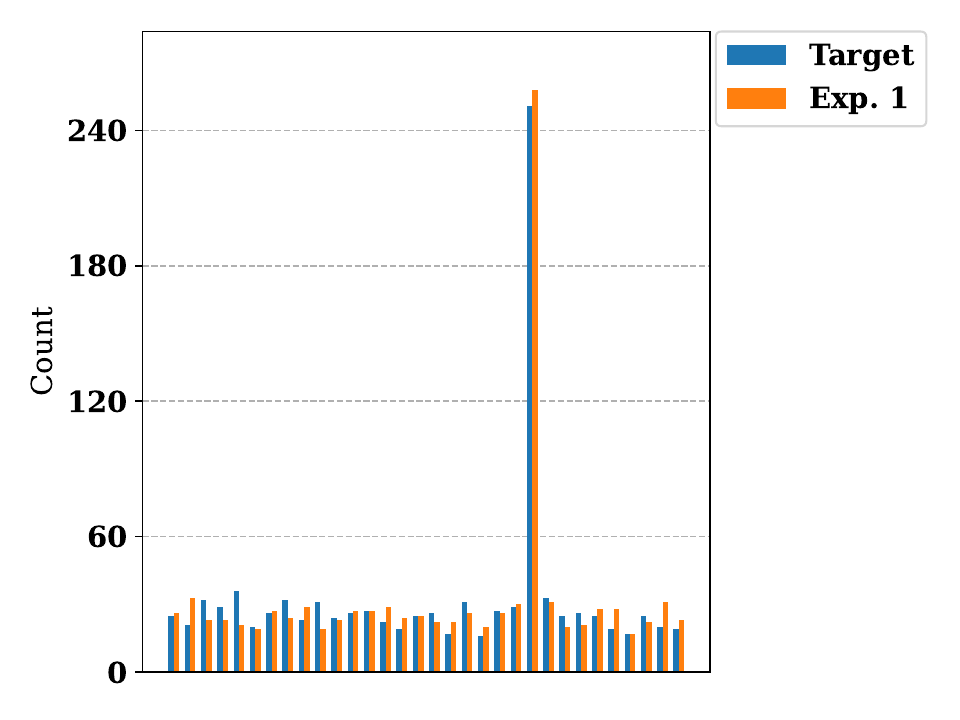}
        \caption{5 qubits distribution where the fidelity weight in the fitness is set to 2.}
        \label{fig:5_qubit_counts_weight2}
    \end{subfigure}
    \caption{Resulting distribution after executing the optimized circuit for the fitness function that reduces the number of CX gates. Each plot is the result of a single experiment.}
    \label{fig:5_qubit_distribution_comparison}
\end{figure*}

\begin{figure*}[t]
    \centering
    \begin{subfigure}[t]{0.3\textwidth}
        \includegraphics[scale=0.30]{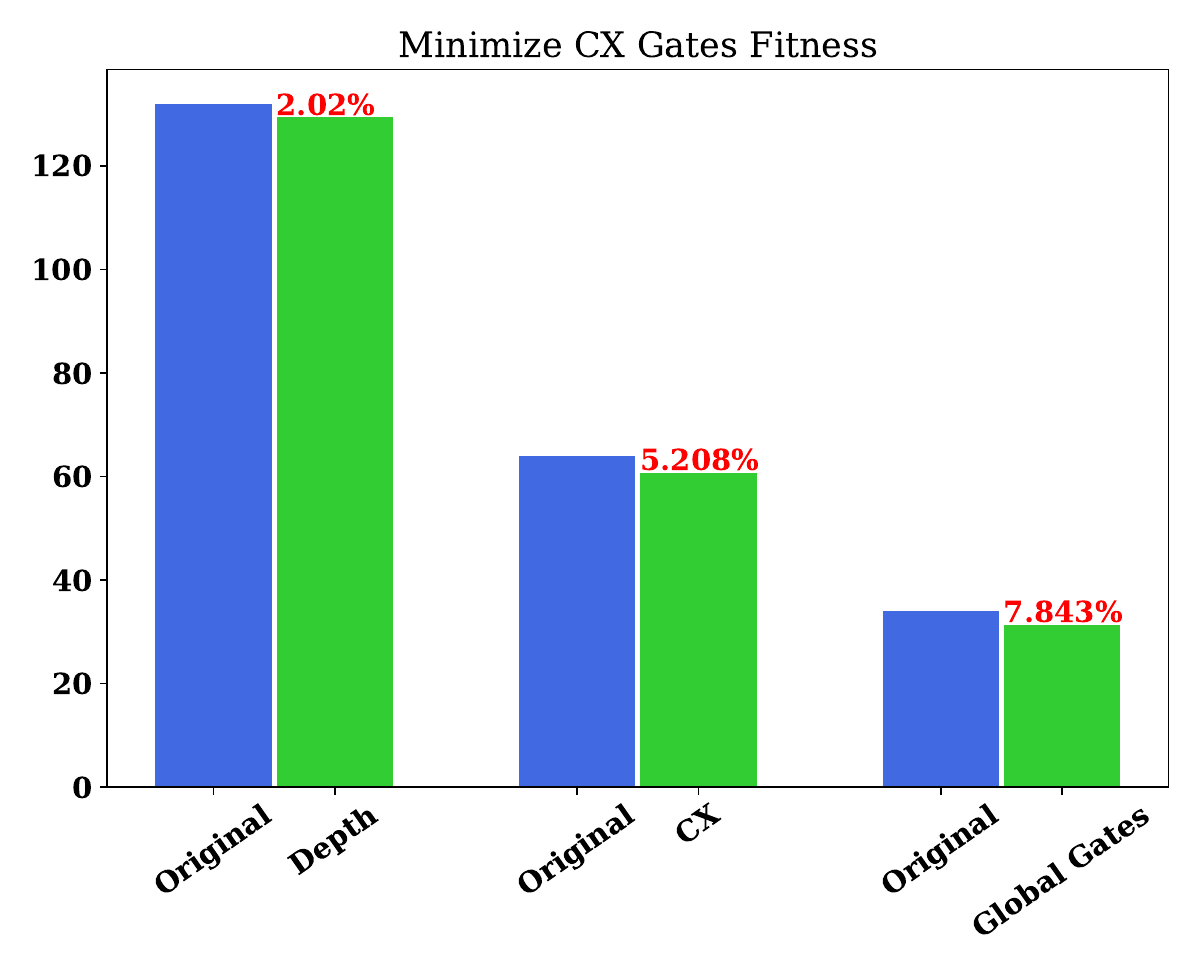}
        \caption{Results of the fitness to reduce the CX gates with fidelity weight of 2 for the 5 qubit experiment.}
        \label{fig:results_reduce_cx_5_qubit_weight2}
    \end{subfigure}
    \hfill
    \begin{subfigure}[t]{0.3\textwidth}
        \includegraphics[scale=0.30]{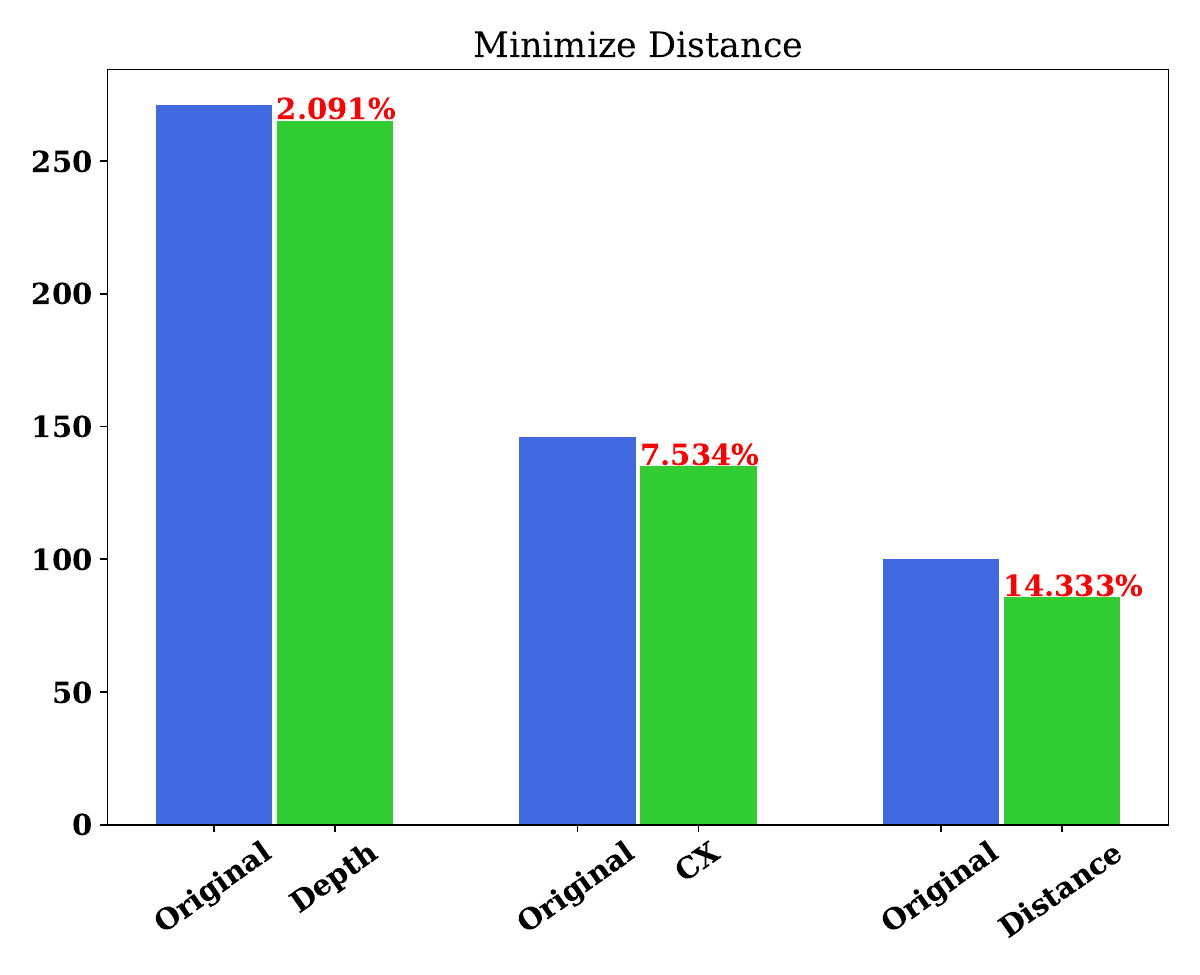}
        \caption{Results of the experiments where a circuit with 6 qubits is optimized for execution on a network where the initial partitioning is given.}
        \label{fig:results_6_qubtits_network}
    \end{subfigure}
    \hfill
    \begin{subfigure}[t]{0.3\textwidth}
        \includegraphics[scale=0.30]{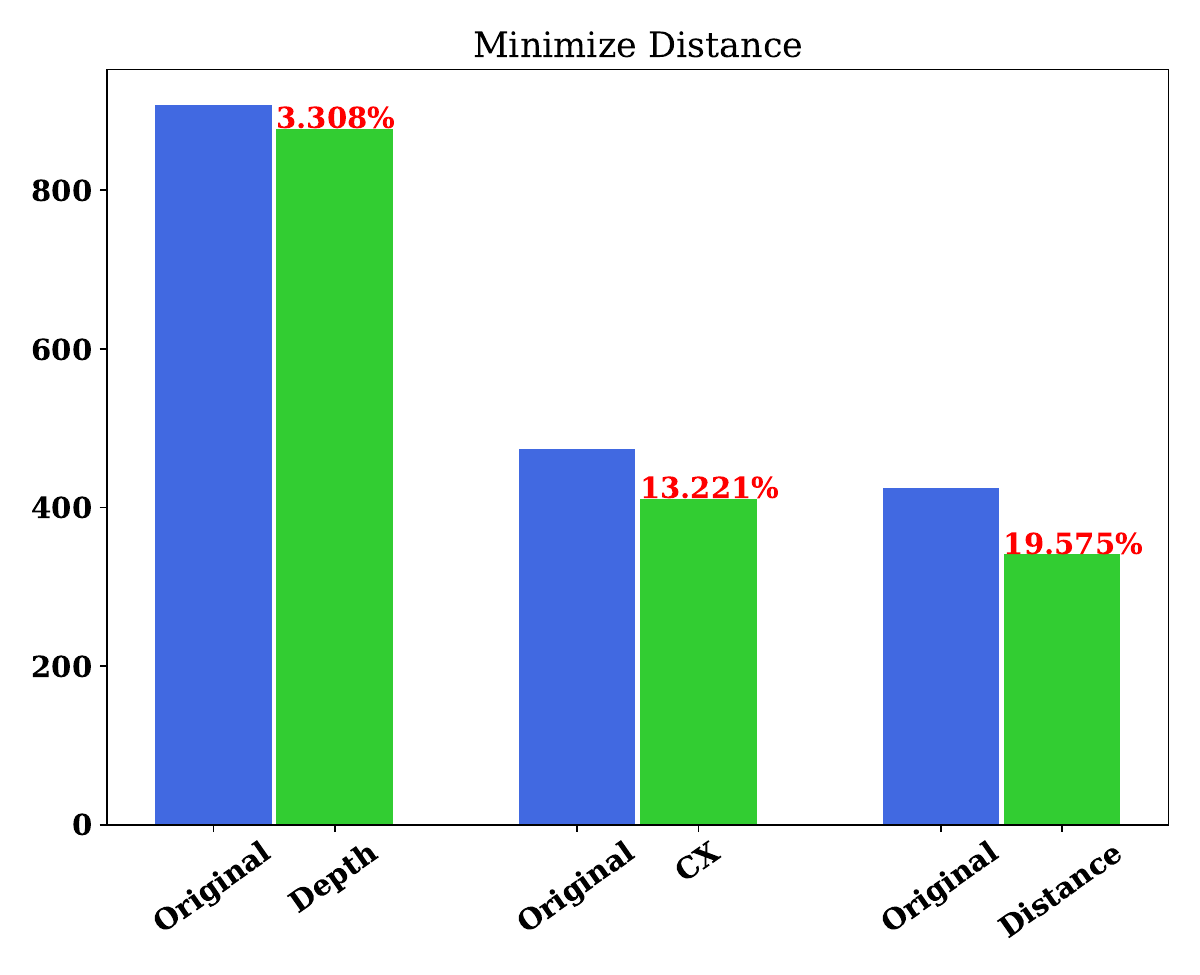}
        \caption{Results of the experiments where a circuit with 8 qubits is optimized for execution on a network where the initial partitioning is given.}
        \label{fig:results_8_qubtits}
    \end{subfigure}
    \caption{Results with 5 qubits and fidelity weight of 2 for the minimize CX gates fitness. Plots (b) and (c) where distance between QPUs is used as fitness.}
\end{figure*}

\section{Results}\label{sec:results}
We present and discuss the experimental results for each fitness function. Note that for all the results in the plots, the correct solution could still be successfully extracted from the resulting state after optimization. Fig. \ref{fig:results_reduce_depth} shows the results of the fitness function that aims to reduce the circuit's depth. Note that in the plots, original (shown in blue) refers to the circuit before optimization, and green results refer to the circuit after optimization. Plots depict the aggregated mean over all seeds. For all circuits, the depth is significantly reduced, i.e., 36\%, 32\% and 27\% for circuits with 4, 5, and 6 qubits respectively. In terms of communication cost minimization, i.e., the reduction of required global gates, the best results were achieved in the 5 qubit circuit experiment with over 26\% reduction. Experiments with 4 and 6 qubits achieved a reduction of 8\% and 5\% respectively. 
Fig. \ref{fig:results_reduce_cx} depicts the results of the fitness function that minimizes the overall number of CX gates. The algorithm is successful in reducing the number of CX gates in all instances, with a rate between 39\% (6 qubits) to over 83\% (5 qubits). While the optimized version of the 5 qubit circuit is still able to produce the desired state, its resulting distribution varies noticeably to its original, as can be seen in Fig. \ref{fig:5_qubit_counts_weight1}. Executing this circuit under noisy conditions may fail at yielding the correct result. Compare this to Fig. \ref{fig:5_qubit_counts_weight2} where the fidelity was given a weight factor of 2 in the fitness function. Here, the distribution is closer to its original circuit; however, the algorithm is not able to reduce the metrics as well as before as shown in Fig. \ref{fig:results_reduce_cx_5_qubit_weight2}. This illustrates the importance of the fidelity weight factor in the fitness function.
The results of experiments in which the number of global gates is directly minimized are shown in Fig. \ref{fig:results_reduce_gg}. Here an improvement of 60.41\%, 15.741\%  and 33.836\% for 4, 5, and 6 qubits respectively, was achieved. 
Figures \ref{fig:results_6_qubtits_network} and \ref{fig:results_8_qubtits} show the results of the experiments in which a circuit is optimized for execution on a specified network topology where the partitioning of the qubits is given. In the 6 qubit experiment, a 3 node network where each node is connected to every other node was used while a grid network consisting of 4 nodes was used for the 8 qubit case. A fidelity weight $\alpha$ of 2 and 3 was used for networks with 3 and 4 nodes respectively. In the 6 qubit experiment, a reduction of 14.33\% was achieved. In the 8 qubit experiment, a reduction of the communication cost of 19.575\% was achieved. Note that here we do not simply compare the global gates, but the distance, i.e, number of hops between QPUs. 

\section{Discussion}
The mean fidelity of the respective experiments is listed in Table \ref{tab:fidelity}. Although the fidelity is not perfect, i.e., less than 1, the optimized circuits were still able to prepare a state similar enough to the original so that the correct solution could still be extracted. However, it is crucial to point out that the circuits are not necessarily functionally equivalent, that is, the underlying unitary may be quite different, even if the circuits prepare an identical state from the given initial conditions. This is an important fact to keep in mind when selecting the appropriate technique for a given task. For the purposes of this paper, the above fitness functions are sufficient; however, when integrating the proposed pre-processing scheme into, for example, a DQC compilation stack alternative fitness functions may be required, although this requires further work. Also, running the experiments for more generations or performing an extensive hyperparamter tuning may yield even better performance. The current configuration suffices as a demonstration of the validity and effectiveness of the approach, however, it by no means is necessarily the limit of what is possible. Furthermore, it depends on the use-case on how one wants to optimize a given circuit. For example, in a software-stack for a quantum network a reduction of, for example, 10\% may already be sufficient whereas in other scenarios more is required. The exact ramifications are subject to further use-case driven research.

\begin{table}[tb]
    \centering
    \caption{The mean fidelity of the best circuits after optimization. The header shows the fitness function. ``Minimize GG'' refers to the fitness function that minimizes the number of global gates directly and ``Minimize Distance'' refers to the experiments where a network topology is used to determine the distance in the fitness calculation.}
    \label{tab:fidelity}
    \begin{tabular}{l@{\hspace{0.3cm}}l@{\hspace{0.3cm}}l@{\hspace{0.3cm}}l@{\hspace{0.3cm}}l}
        \toprule
        & Min. GG & Min. CX & Min. Depth & Min. Distance \\ \midrule
         4 Qubits & 0.972087 & 0.91407 & 0.915459 & - \\ \midrule
        5 Qubits & 0.968894 & 0.901349 & 0.94335 & - \\ \midrule
        6 Qubits & 0.966656 & 0.949755 & 0.94693 & 0.990059 \\ \midrule
        8 Qubits & - & - & - & 0.984862 \\ 
        \bottomrule
    \end{tabular}  
\end{table}

\section{Conclusion}\label{sec:conclusion}
While DQC is a viable approach to scaling quantum computing, it also has its drawbacks and challenges. Communication overhead in the form of quantum teleportation or remote gate execution is one of the arising problems in this new paradigm. We applied an EA to optimize a given circuit so that the number of required non-local gates are reduced. In all experiments, the optimized circuit was able to reduce the communication overhead up to 30\% while reducing the fidelity only slightly (mean above 0.9 in all instances) and, more importantly, the circuits were still able to produce the desired state and extract the correct solutions. We furthermore ran experiments in which a circuit is optimized to fit a given qubit assignment in a network. Here, the EA was also able to reduce the communication cost by 14\% and 19\% for 6 and 8 qubit circuits respectively while retaining a fidelity of above 0.98. However, running the algorithm for more generations or performing a comprehensive hyperparameter optimization may result in even better performance. Future work could investigate how the approach can be integrated into a compilation stack for DQC and how the approach can be adjusted for other use cases that require functional equivalent circuits.

\bibliographystyle{IEEEtran}
\bibliography{IEEEabrv,bibliography}

\end{document}